\newcommand{\logg}{$\log~g$}

\documentclass{aa}
\usepackage{graphicx}

\begin{document}

   \title{Some Aspects of the calculation of Balmer lines in the sun and 
    stars}

\author{C. R. Cowley\inst{1}
\and
           F. Castelli\inst{2}
 }

\institute{Astronomy Department, University of Michigan, Ann Arbor,
 MI 48109-1090 USA,
             \email{cowley@astro.lsa.umich.edu}
\and
CNR-IAS and Osservatorio Astronomico 
di Trieste, Via G.B. Tiepolo 11, I-34131 Trieste, Italy,
            \email{castelli@ts.astro.it}
 }
\offprints{C. Cowley}

\date{Received ; accepted }

\titlerunning{Balmer lines in the sun and stars}

\abstract{We compare the results of Balmer-line calculations using 
recent theory and improved computational algorithms with those 
from the widely-used SYNTHE and BALMER9 routines.  The resulting
profiles are mostly indistinguishable. 
Good fits to the normalized solar Balmer lines H$\alpha$ through H$\delta$
are obtained (apart from the cores) using  
the recent unified-broadening calculations by Barklem and his
coworkers provided that some adjustment for the continuum is performed.
We discuss a surprising linearity with temperature
of the Balmer line profiles in dwarfs.   
}

\maketitle

\section{Introduction}

\begin{figure}
\resizebox{9cm }{!}
{\includegraphics[0., 0.][450., 450.]{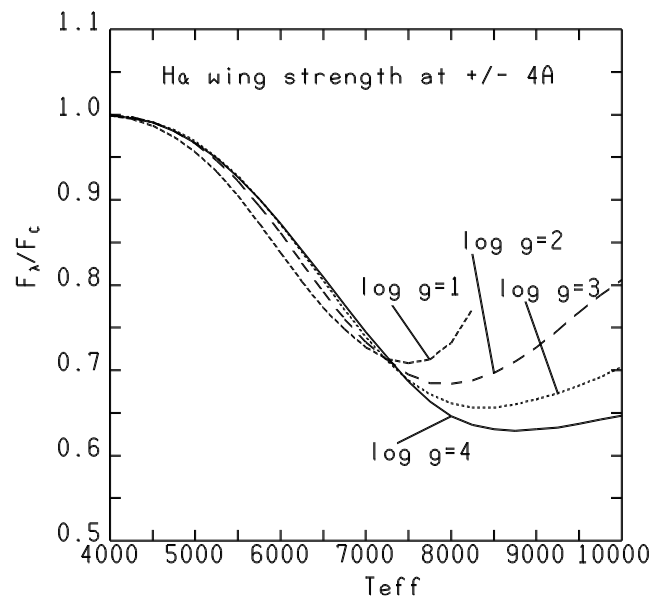}}
      \caption{H$\alpha$ wing strength vs. $T_{\rm eff}$ for
       several values of \logg. The profiles are taken from the BP00K2NOVER grid
       available  in http://kurucz.harvard.edu}
      \label{wing4.ps}
   \end{figure}

Balmer line strengths are highly sensitive to the temperature 
in cool stars because of the 10.2 eV excitation of the $n=2$ 
level from which they arise.  Fig. 151 from Uns\"{o}ld's (1955)
classic text illustrates this for H$\gamma$ equivalent widths.
We show the effect in a different way in Fig.~\ref{wing4.ps},
based on more recent line-broadening theory. 
The figure is for points on the H$\alpha$
profile 4~\AA\ from the line center, but is characteristic of
much of the line profile.   


An extensive investigation of Balmer lines in cool dwarfs 
(Fuhrmann, Axer, \& Gehren 1993; Fuhrmann, Axer, \& Gehren 1994) 
concluded these lines 
provide a more consistent guide to effective temperatures than 
broad-band colors or $b - y$.    Nevertheless, Balmer line 
profiles are not regularly used to fix the effective temperature of cool 
stars.  The reasons for this are numerous, but have not been
explicitly addressed.  Some insight may be gained from the papers 
by Van't Veer-Menneret \& M\'{e}gessier (1996) or Castelli, 
Gratton \& Kurucz (1997, henceforth, CGK).  A recent paper which
does discuss use of H$\alpha$ in the determination of effective
temperatures is by Peterson, Dorman \& Rood (2001). 
In addition to the uncertainties in placing the continuum
level, uncertainties, both in the theory of stellar atmospheres 
($l/H$, convection) and line formation remain unresolved.

The absorption coefficient of neutral hydrogen  takes into account 
the effects due to the natural absorption (natural broadening),
the velocity of the absorbing hydrogen atoms 
(thermal Doppler and microturbulent
broadening), the interactions with charged perturbers 
(linear Stark broadening),  with neutral perturbers different from
hydrogen (van der Waals broadening), and with neutral hydrogen
perturbers (resonance and van der Waals broadening). 
Each effect is represented by a profile and the
total effect requires a convolution.
Thermal Doppler and microturbulent broadenings are 
described by gaussian functions
while natural, resonance, and van der Waals broadenings 
have  Lorentz profiles.   These two profiles are combined into 
a Voigt function.  The convolution of the Voigt profile with the
Stark profile or Stark plus thermal Doppler effect then gives
the  total absorption profile.
 
Most of the damping constants and Stark profiles are computed from 
complex theories based on several approximations, while the complete 
convolution of all the above profiles 
is a very time consuming algorithm.

In this paper we describe our attempts to evaluate several aspects
of the calculations of Balmer line profiles.

\section {Stark profiles}

Most work on 
stellar atmospheres makes use of codes provided by Kurucz 
(http://kurucz.harvard.edu). For computing hydrogen lines the codes 
are either BALMER9 (Kurucz, 1993a) which produces
profiles for H$_{\alpha}$, H$_{\beta}$, H$_{\gamma}$,
and H$_{\delta}$ or the SYNTHE code (Kurucz, 1993b) which produces profiles for
any hydrogen line. In the first case  Stark profiles  
are interpolated in the Vidal, Cooper, \& Smith (1973, henceforth VCS)
tables, while in the second case the Stark profiles are based on the 
quasi-static Griem theory with parameters adjusted in such a way
that profiles from Griem theory fit  the VCS profiles of the 
first members of the Lyman and Balmer series.
 
Only the most recent work on the Balmer lines 
(e.g. Barklem, Piskunov \& O'Mara 2000, henceforth, BPO)
has included the new Stark profiles of Chantal Stehl\'{e}
(henceforth CS) and her coworkers.  They are 
available from a link on her website:
http://dasgal.obspm.fr/~stehle/.
A recent reference is Stehl\'{e} \& Hutcheon (1999).

\begin{figure}
\resizebox{8.0cm }{!}
{\includegraphics[angle=-90]{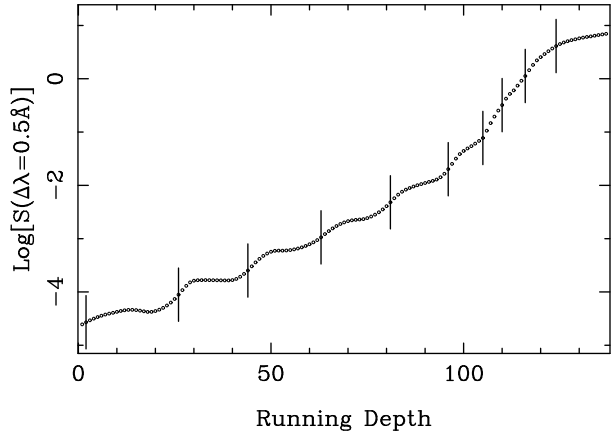}}
      \caption{Normalized Stark width at $\Delta\lambda = 0.5$\AA
               \ for H$\alpha$ vs. 137 depths in an 
               Holweger-M\"uller (1974) solar model.
               Each depth step is 0.05 in $\log(\tau_{\lambda 5000})$.  
               The vertical lines mark depths corresponding to 
               boundaries of
               the tables giving $S(\alpha)$ for a fixed value of 
               the electron density.
              }
       \label{waves2.ps}
   \end{figure}

A problem arises when a given Stark profile is interpolated 
either in the VCS  or in the CS tables by using the
interpolation method taken from the BALMER9 code. This is
 a bilinear interpolation in $\log(T)$ and 
$\log(N_e)$, followed by a linear interpolation in the 
parameter $\Delta\alpha = \Delta\lambda[{\rm \AA}]/F^0$.  Here, 
$F^0$ is the normal field strength in Gaussian cgs units, $F^0 
= 1.25 N_e^{2/3}$, so the interpolation in $\Delta\alpha$ is 
not independent of the previous one which involves the
electron density $N_e$. We 
find this introduces a small error that shows up as an oscillation 
in a plot of the Stark profile $S(\Delta\alpha)$ vs depth in 
the solar atmosphere for a small range of displacements from 
the line center as shown in Fig.~\ref{waves2.ps}.

We were able to remove the oscillations by 
rewriting the CS tables with $\Delta\lambda$ as the third 
(independent) variable, and using essentially the same 
interpolation scheme as BALMER9.  Fortunately, it has resulted that 
the improved interpolation leads to no perceptible changes
in the resulting line profiles.

\section{Convolution of profiles and microturbulence}

Neither the BALMER9 code nor the SYNTHE code 
perform profile convolutions, but all the profiles are simply
added. In the BALMER9 code, for separations larger than 0.2~\AA\ from
the line center, a Lorentz profile
(representing the natural broadening and the resonance broadening)
is added linearly to the  Stark-thermal Doppler profile 
interpolated in the VCS tables. 
For separations smaller than 0.2~\AA\ no Lorentz profile was considered. 

In the SYNTHE code, the Doppler profile,
the Stark profile, and the Lorentz profile (for 
natural broadening, resonance broadening, and van der Waals broadening
from He~I and H$_{2}$) are still summed together.    
The very inner core is that of the profile (Doppler, Stark, or Lorentz)
with the largest full width at half maximum FWHH.

This method due to Peterson (1993), which we shall call the PK approximation,
 would be rigorously true for the 
wings of two Lorentzians. Since the wing-dependence of the 
Stark profile differs from that of a Lorentzian only by 
$\sqrt(\Delta\lambda)$, one might expect the approximation to 
be good, as we verified that it is.

\begin{table*}
\caption{Models used for H$\alpha$ tests}
\begin{center}
\begin{tabular}{r c l r}\hline
$T_e$(K)&\logg& $\xi_t$(km~s$^{-1}$)  & Comment \\ \hline

4500  &  1.5 &   3.0   &  solar abundances  \\      
4760  &  1.3 &   2.3   &  CS22892-052 (cf. Sneden et al. (1996)) \\
5770  &  4.4 &   1.0   &  sun             \\
8000  &  3.5 &   2.0   &  like cool Ap or Am \\
8000  &  1.5 &  12.0   &  test of large $\xi_t$ \\
12000 &  3.0 &   2.0   &  hot star \\  \hline

\end{tabular}
\end{center}
\end{table*}

\begin{figure}
\resizebox{8.0cm }{!}
{\includegraphics[angle=-90]{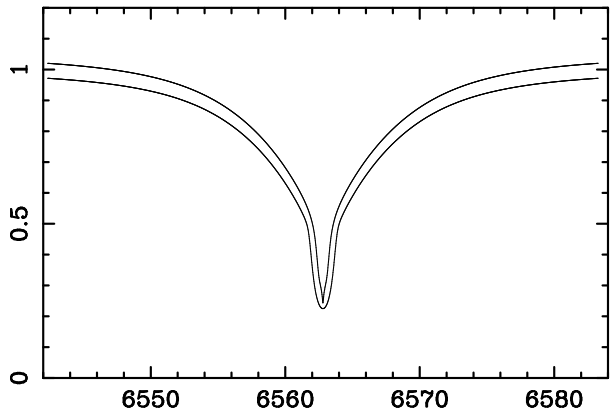}}
      \caption{H$\alpha$ profiles for a model with $T_{\rm eff} = 8000K$, 
       \logg=1.5. The lower curve is for a CSII calculation
     with an assumed microturbulence $\xi_{\rm t} = 12$ km~s$^{-1}$. 
     The upper curve, displaced upward for purposes of illustration,
     was made using BALMER9, 
     the older interpolation scheme for VCS tables, and the PK
     approximation.  There is no perceptible difference in the two
     profiles beyond the line core.} 
      \label{2plot.ps}
   \end{figure}

Replacing the sum of the Stark and Lorentz profile in BALMER9
by a convolution  takes a large amount of computing time in that 
the $\Delta\lambda$ step of
the convolution has to be very small (less than
0.001 \AA) in order to account for the narrow full width at half
maximum FWHM  of the Lorentz profile.
This problem can be overcome by including a microturbulent
velocity $\xi_t$ in the computations.

Both the VCS and CS tables include thermal Doppler, 
but not microturbulent broadening.  
The BALMER9 code makes no provision for the
inclusion of microturbulence in the line profiles 
owing to the sum of
the  Stark-thermal-Doppler profile, interpolated in the VCS tables, with
the Lorentz profile.
The SYNTHE code does allow for a microturbulence in that it adds
the Stark profile to a Doppler-microturbulence gaussian profile. 

The only way to 
rigorously include all broadening mechanisms is to do a convolution of the 
Stark-thermal Doppler profile, interpolated in the VCS or CS tables, with a 
profile which includes both the Lorentz broadening and 
turbulent motions.  If we assume a Gaussian distribution of microturbulent 
velocities, the VCS or CS profiles need to be convolved with a Voigt
profile.

To check BALMER9 and SYNTHE profiles we did calculations using 
the new CS profiles
with improved interpolation, and a  full convolution including a
microturbulent velocity. We shall refer to such profiles 
and to the corresponding code with the 
abbreviation CSII (Convolution, Stehle, improved interpolation). 
Table 1 shows models parameters for which we made 
calculations of an 
H$\alpha$ profile in order to test the effects of the various 
approximations and improvements mentioned above.  All models
were generated with the ATLAS9 code (Kurucz, 1993a).
  Solar abundances were assumed for all
but CS22892-052, for which abundances were chosen to roughly
match those of Sneden et al. (1996).

We find, with one exception, that the BALMER9 profiles
computed with no convolutions and no microturbulent
velocity are in excellent agreement with CSII calculations.
The only exception occurs for the supersonic microturbulent
velocity $\xi_t$=12~km~s$^{-1}$. In this case the line core
of the profile computed for  $\xi_t$=12~km~s$^{-1}$ is
larger than that computed without microturbulence, as is  
shown in Fig.~\ref{2plot.ps}. 
However, the H$_{\alpha}$ profile computed by SYNTHE 
with no convolutions, but 
by assuming $\xi_t$=12~km~s$^{-1}$  agrees well with 
the CSII profile.  

 The effect of a 
microturbulent velocity $\xi_t$ will be small until $\xi_t$ 
approaches the sound speed.  
It is not surprising, therefore, that the {\it only} case we
have found where plots of H$\alpha$ obtained using  BALMER9 with
the PK approximation  and 
CSII differed significantly is that for $\xi_t$  of 
the order of the sound speed.  Even in this situation, only 
the deepest parts of the core were affected.  The line wings 
still matched beautifully.
 
The calculations of Fuhrmann et al. (1993;1994) included Lorentz broadening by a
full convolution, while BPO used the PK approximation.
The above comparisons led us to conclude
that any differences between their results and other calculations (e.g.
CGK or Gardiner, Kupka \& Smalley 1999) cannot be
attributed to the PK approximation or to different Stark
profiles (VCS or CS) -- the immediate line core excepted.

\section {Broadening of the hydrogen lines by collisions with H~I atoms}

The BALMER9 and SYNTHE codes allow for the broadening of
the hydrogen lines due to the collisions with other 
neutral H~I atoms through
the resonance broadening based on the Ali \& Griem theory (1965,1967).
Actually the van der Waals effect due to H~I should also be included, 
but it can not be simply added to the resonance broadening 
(Lortet \& Roueff, 1969) and therefore it was always neglected in 
the hydrogen profile calculations. Only recently BPO (Barklem et al. 2000)
presented an unified theory of
the H~I-H~I collisions in the stellar atmospheres.
The differences in Balmer profiles computed with only 
resonance broadening and with both  resonance
and van der Waals broadenings are fully discussed in
BPO.

We have included in our hydrogen synthetic spectra (BALMER9, SYNTHE and 
CSII) the BPO broadening.
The line half half-width HWHM per unit hydrogen atom density $w/N(H)$ is
computed according to Anstee \& O'Mara (1995):

$$w/N(H)=(4/\pi)^{\alpha/2} \Gamma(2-\alpha/2) v \sigma(v_{0}) (v/v_{0})^{-\alpha}$$
where the cross-section $\sigma$ and the velocity parameter $\alpha$ 
for H$\alpha$, H$\beta$, and H$\gamma$ were taken from Table~3 in BPO. 
Furthermore, we recall that
$v=(8RT/\pi\mu)^{1/2}$, where $\mu$ is the reduced mass 
for two hydrogen atoms,
and $v_{0}$ is the velocity $v$ for 10$^{6}$ cm~s$^{-1}$. 
The value of the $\Gamma$ function is 0.901903 for H$_{\alpha}$, 0.92437 for
H$_{\beta}$ and  0.93407 for H$_{\gamma}$.

In the CSII code, HWHM was computed in according to BPO for each 
given temperature of the atmospheric layers. For H$_{\delta}$
the broadening by neutrals was obtained by extrapolating BPO's Table 3, but
the profile is dominated by Stark broadening, and is nearly 
independent of the broadening by neutrals.  
In BALMER9 and in SYNTHE,  HWHM 
was obtained for each temperature of the atmospheric layers
from  a function HWHM=HWHM$_{0}$~(T/10000)$^{y}$ 
where  HWHM$_{0}$
is the value of HWHM for T=10000~K and {\it y} was derived  
from the best fit of the above function to the HWHM,T points  
for T ranging from 2000~K to 11500~K at steps of 500~K
(Fig.~3 in BPO). The parameter {\it y} is
0.15 for H$_{\alpha}$, 0.275 for H$_{\beta}$, and 0.30 for H$_{\gamma}$.   

\section{Balmer profiles from the Holweger-M\"uller solar model}

\subsection{The solar HM Model}

For the calculation of the solar Balmer profiles we adopted the
Holweger-M\"uller model (1974, henceforth, HM) to avoid additional
complications from various solar models, already discussed, for example,
by CGK. 
We started from the HM T-$\tau_{5000}$ relation given for 29
layers, and extrapolated-interpolated to suit the depth
ranges used by our respective codes. 

There are  differences in the optical depth 
coverage of the  Michigan and Trieste codes.
In the first case, the  T-$\tau_{5000}$ relation was 
interpolated-extrapolated to 135 layers, while in the second case
it was interpolated for 50 layers before using it in the Kurucz codes.
While the Michigan code performs integrations directly in terms
of $\log(\tau_{5000})$,
the use of the Kurucz codes 
requires a conversion from  the $\tau_{5000}$ depth scale
to a RHOX (or $\int\rho dx$) depth scale,
where $\rho$ is the density of the stellar gas
and $x$ is the geometrical height in
the atmosphere. The conversion was obtained by computing the
continuous opacity $\kappa_{5000}$ at $\lambda$=5000~\AA\ by means of the
ATM code from Holweger, Steffen \& Steenbock (1992, private communication) 
and by deriving RHOX  from the relation 
$d\tau_{5000}$ $=$ $\kappa_{5000}\rho~dx$.
The original HM
model was made more than a quarter of a century ago.  Since that
time, abundances and
the continuous opacity routines have been modified,
presumably for the better.  This means that the current relation between
$\tau_{5000}$ and $\tau_{\rm Rosseland}$ is no longer the same
as in the HM paper.  The latter is inconsistent with the RHOX scale 
of the modern Kurucz codes.

We adopted as solar abundances  the meteoritic values from
Grevesse \& Sauval (1998) and a constant
microturbulent velocity $\xi$=1~km~s$^{-1}$.

The HM model used in the Kurucz codes is given in the
Appendix~A.
\subsection{Predictions from the HM model}

\begin{figure*}
\resizebox{14.0cm }{!}
{\includegraphics[0.,0.][650.,550.]{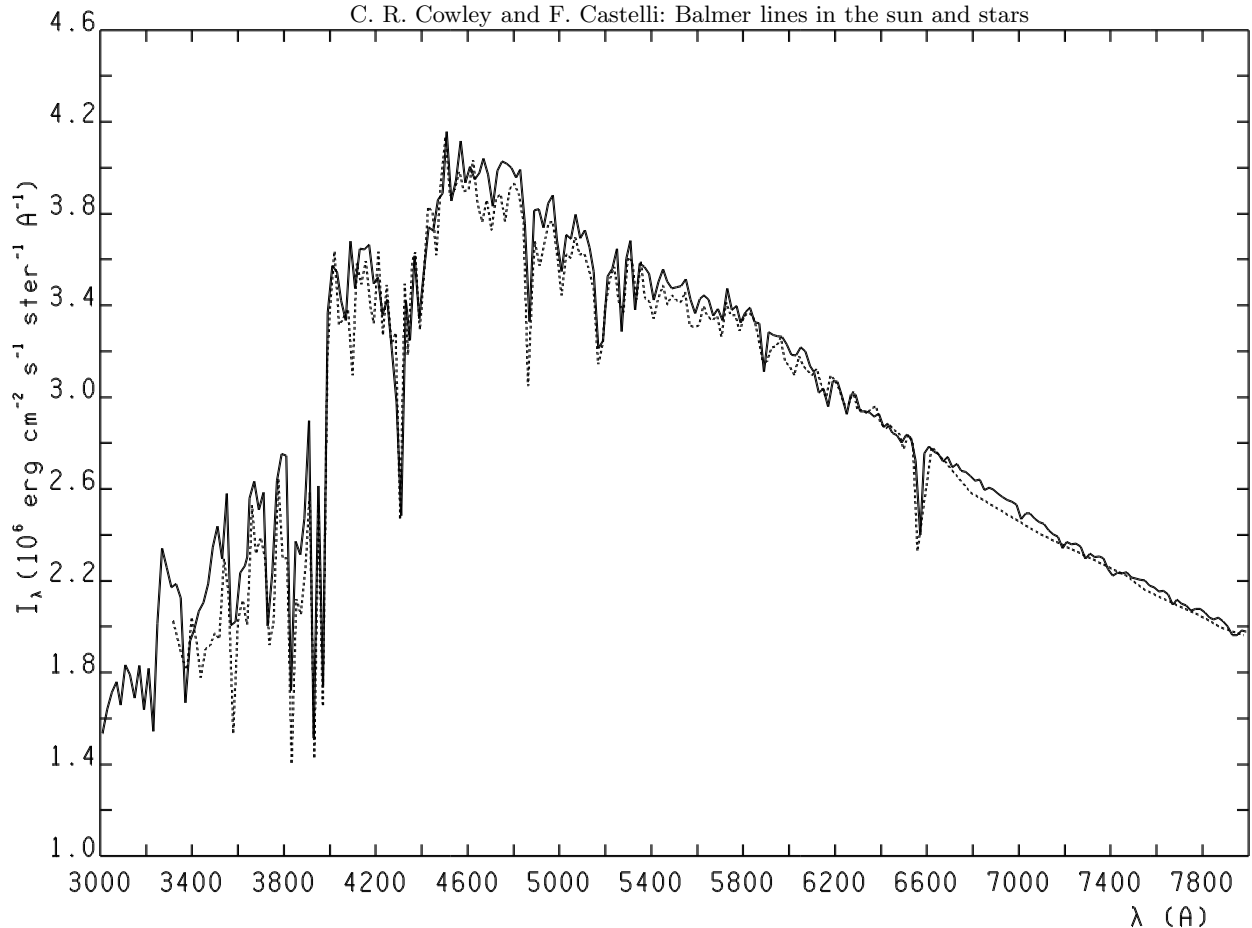}}
      \caption{Comparison of the solar intensity from 
the center of the sun predicted by the HM model (full line)
with the observations from Neckel \& Labs (1984) (dashed line).  The line opacity
in this low-resolution calculation is entirely from the ODFs.}
      \label{psunichm}
   \end{figure*}

For clarity, we first list several categories of opacity relevant
to the current problems:
\begin{enumerate}
\item Standard continuous opacity: bound-free and free-free transitions
in various atoms and ions, Rayleigh and Thomson scattering.
These are implemented in most currently-used model atmosphere
and spectrum synthesis codes.

\item TOPBASE opacities (Seaton et al. 1992).   These opacities
have not yet been widely implemented in current atmosphere codes,
so the impact of this important work remains to be seen.

\item Line opacity due to transitions between tabulated atomic
energy levels.  Some of these lines are predicted, in the sense
that they have not been observed on the laboratory, but all 
relevant {\it levels}  have been located, typically to a fraction
of a wavenumber from observed lines.  We shall call these {\it
classified lines.}
We distinguish two categories:

\begin{enumerate}
  \item Stronger lines, which contribute 1\% or more to the continuous
opacity at the central wavelength for point in a model atmosphere.  
  \item Weaker lines, for which the above criterion is not met.
\end{enumerate}
\item Line opacity due to transitions involving one and sometimes two
levels whose locations are predicted by an atomic structure code.
Wavelengths for these lines may be uncertain by 10 or more
angstroms.  A sizable fraction of these lines involve levels above
the first ionization limit, and the levels are therefore subject
to autoionization.  We shall  refer to these as {\it unclassified lines.}
Many of these lines may have been observed in laboratory experiments. 
Again, we list two categories:
  \begin{enumerate}
   \item Stronger lines.  In certain chemically peculiar stars, we know
there must be many such lines because we are unable to identify a
large fraction of the measurable stellar lines.  There are also many
unidentified lines in the solar spectrum, though they are usually weaker
than a few tens of milliangstroms, and typically increase in number 
to the violet. 
  \item Weaker lines connecting predicted levels.
\end{enumerate}
\item ``Missing'' opacity.  Calculations of the solar continuum
using only standard continuous opacity (No. 1 above) predict values
significantly higher than the ``observed''
continuum.  The disparity increases toward the violet
(see discussion below).
\end{enumerate}

 Fig.~\ref{psunichm} compares the solar intensity I$_{\lambda}$(0) 
 from the center of the Sun
measured by Neckel \& Labs (1984) with I$_{\lambda}$(0) predicted using 
the continuous and line opacities from Kurucz (1993c) and the HM model
given in Appendix~A. The line opacity is treated with the opacity 
distribution functions (ODF), which include both classified and unclassified lines.
When such a line opacity is considered in the calculations, 
a rather good agreement of the low resolution observations with the
low resolution predictions is obtained, indicating that much of the missing 
opacity could be due to line absorption.
Because the ODFs involve averages over wavelength intervals of the
order of 20~\AA\ in the 3300-6400~\AA\ region and larger for
$\lambda$$>$6400~\AA, we
refer to the calculation of Fig.~\ref{psunichm} as a
low-resolution synthesis.
The nature of the missing opacity is somewhat controversial,
and will not be argued here.  A recent reference, with citations
to earlier discussion, is Peterson et al. (2001).

\begin{figure}
\resizebox{9.5cm}{!}
{\includegraphics[bb=0. 0. 700. 450.]{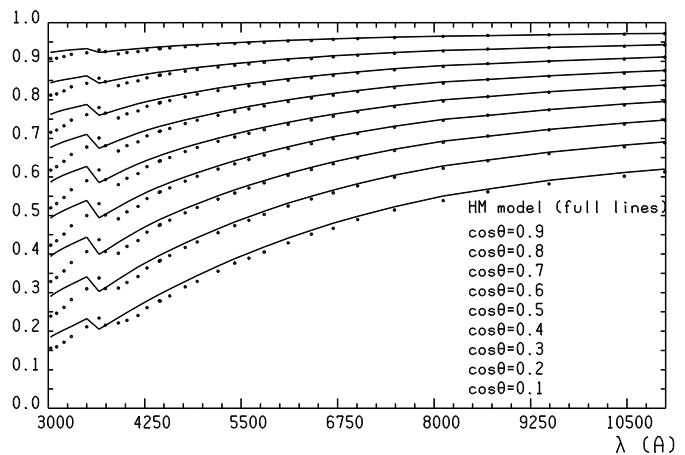}}

      \caption{ Comparison between observed (points) and computed
(full line) solar limb-darkening curves 
I$_{\lambda}$(cos$\theta$)/I$_{\lambda}$(0). Observations are from
Neckel \& Labs (1994) and computed curves are based on the HM model
     }
      \label{hmld}
   \end{figure}

Limb darkening predictions from the HM model are compared in
Fig.~\ref{hmld} with those from Neckel \& Labs (1994).
In this case, opacity from lines
is not included in the computations in accordance with 
the assumption of Neckel \& Labs (1994) of observations made
at wavelengths free from lines contaminating the continuum.
The departure of the computations from the observations
in the violet can be explained with the poor chance to have
regions free from lines in this part of the solar spectrum.
Except for the violet wavelengths,  the agreement is satisfactory.

\subsection{The Balmer profiles in absolute intensity}

\begin{figure}
\resizebox{8.0cm }{!}
{\includegraphics[bb= 0. 0. 600. 1450.]{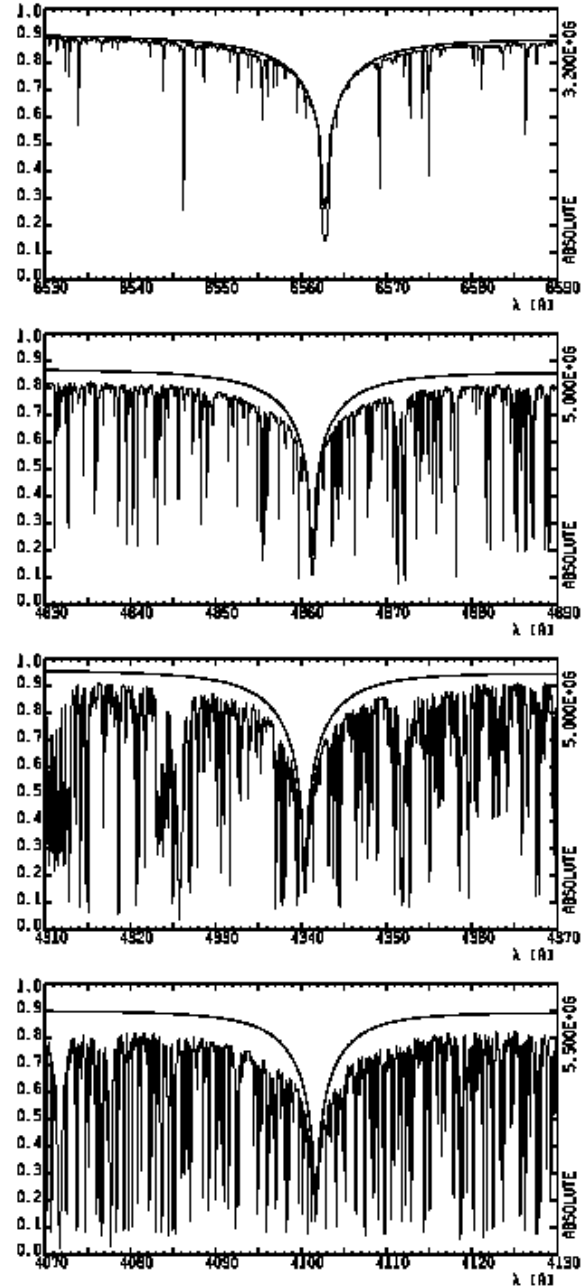}}
\caption{Comparison between observed and computed Balmer profiles
in absolute I$_{\lambda}$(0) units.  The left scale gives intensities
relative to the following values, all in units of
$10^6$ ~erg~cm$^{-2}$~s$^{-1}$~stear$^{-1}$~\AA$^{-1}$.
For H$_{\alpha}$, the maximum of the uppermost panel is  3.2,
for H$\beta$ and H$\gamma$ the maxima are both 5.0, and
for $H\delta$, the maximum of the panel is 5.5.}
\label{habs}
\end{figure}

Fig.~\ref{habs} shows the Balmer profiles for the disk center
in absolute intensity.  We have adopted the Kitt Peak observations 
available at the Hamburg site
(ftp.hs.uni-hamburg.de; pub/outgoing\-/FTS-Atlas) and described
 by Neckel (1999, henceforth, KPN).  The files
include absolute intensities, as well
as continuum estimates at each wavelength. 
The resolution of the observations is about 350000. 

The synthetic Balmer profiles were computed with the SYNTHE code
and the HM model. Only the relevant Balmer line and standard
continuous opacity sources were used.
The spectrum was degraded at the observed resolution and it was
broadened by assuming a macroturbulent velocity $\xi_{macro}$=
1.5 km~s$^{-1}$, although Balmer lines are independent of 
instrumental and macroturbulence broadenings of the order of
those here adopted.

Fig.~\ref{habs} indicates that
except for H$_{\alpha}$, the observed and computed profiles can not be
directly compared  in absolute intensity, owing to the different 
levels of the observed and computed spectra.
The differences are less than 1~\% for H$_{\alpha}$,  of the order of
5~\% for H$_{\beta}$,
 4~\% for H$_{\gamma}$, and  8~\% for H$_{\delta}$.
The discrepancy does not change if all the classified lines are included
in the computations (see also Fig.~7 in CGK).

%

The agreement shown by Fig.~\ref{psunichm} shortward of 4600~\AA\ appears
better, and apparently contradictory, to that of Fig.~\ref{habs}.
However, both resolution and opacities are different in the two
figures.  In addition to the averages over the 20~\AA\ intervals,
Fig.~\ref{psunichm} includes opacities from unclassified lines. 
When the whole line opacity is averaged over the 
20~\AA\ intervals the detailed spectral differences at each wavelength
are smoothed off. The result is a better agreement of 
the spectra observed and computed at low resolution
than that of the same spectra analyzed at high resolution.

The picture is complicated by absolute measurements by
Burlov-Vasiljev et al. (1995) of
the solar spectral energy distribution. It is higher by about 6\% than that of
Neckel \& Labs (1984)  at H$_{\delta}$, 4\% at H$_{\gamma}$,
2\% at H$_{\beta}$, while it is about 2\% lower at H$_{\alpha}$.

We now turn to a comparison of normalized profiles, in which the
above problems are less obvious, though nevertheless present.

\subsection{The normalized Balmer profiles}

In the current work, one of us (CRC)
attempted new estimates of the continuum for the observed spectrum 
---less as an attempt to
improve on the KPN values, as to gain some insight into the 
uncertainties in this endeavor.  We began with spectral high
points within 10~\AA\ intervals plotted vs. wavelength, and 
smoothed the ``envelope'' by selectively deleting points, in an
obviously subjective way, to achieve an overall smooth plot.
The adopted
points are shown in Table~\ref{tab:cont}, along with those
from KPN.  We make no claim that the current continuum is
superior in any way to that chosen in KPN.  It was simply used
in the Michigan work for normalization purposes.  We employed a 
four-point Lagrange interpolation scheme to normalize observations
between the chosen points.

\begin{table}  
\caption{Solar continuum specific intensity in units of
$10^{15}$ cgs}
\label{tab:cont} 
\begin{tabular}{c c c} \hline
Wavelength  &  This work  & KPN  \\ \hline
3298.973& 0.3235& 0.3231 \\
3355.431& 0.3269& 0.3272 \\
3782.919& 0.4083& 0.4093 \\
4020.705& 0.4589& 0.4591 \\
4279.262& 0.4652& 0.4666 \\
4419.404& 0.4598& 0.4609 \\
4504.079& 0.4540& 0.4545 \\
4861.000& 0.4230& 0.4179 \\
5102.095& 0.3999& 0.3990 \\
5203.252& 0.3906& 0.3902 \\
5801.460& 0.3435& 0.3424 \\
6109.561& 0.3200& 0.3189 \\
6202.178& 0.3146& 0.3144 \\
6409.847& 0.2990& 0.2972 \\
6500.584& 0.2907& 0.2899 \\
6802.324& 0.2660& 0.2663 \\
6850.076& 0.2619& 0.2627 \\
6950.356& 0.2546& 0.2553 \\
6972.875& 0.2536& 0.2540 \\
7000.000& 0.2524& 0.2524 \\ \hline
\end{tabular}
\end{table}

Our independent evaluation of the continuum based on
the points shown in Table~\ref{tab:cont} is in excellent
agreement with KPN, with the exception of the region near
H$\beta$.  The value shown in column 2 for $\lambda$4861
interpolated with the four-point Lagrange formula,
from the surrounding points, is 1.2\% higher than the KPN
continuum.  This region appears depressed for reasons that
are unclear and deserve investigation.   

The continuous specific intensity using the HM model
and Michigan codes matches the interpolated
continuum from Tab.~\ref{tab:cont} at H$\alpha$ to within 1\%.
For H$\beta$ through H$\delta$, the calculated continua 
fall above the measured (as interpolated
in Tab.~\ref{tab:cont}) continua by 2.4,
3.9, and 7.8\% respectively.  These results agree well with those
discussed in the previous section of
the comparison of the observed and computed
absolute intensities. 

If we assume the ``missing opacity'' as cause for these disagreements
as well as for those shown in  Fig.~\ref{habs},
there is at present no obviously correct way to account for it.
For these calculations, we assumed this opacity has the same
depth dependence as standard continuous opacity sources.  We
have simply scaled them by constant factors until the calculated
specific continuous intensities agree with the observed chosen continuum.

\begin{figure}
\resizebox{7.0cm }{!}
{\includegraphics[angle=-90]{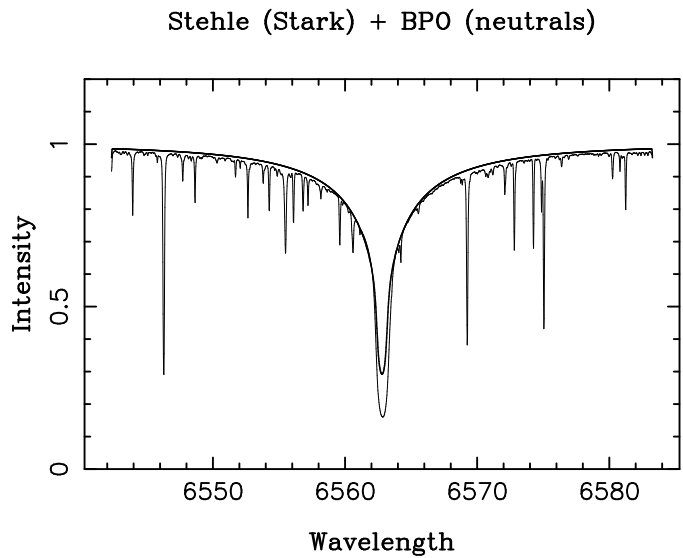}}
      \caption{H$\alpha$ profile for the center of the solar disk
       normalized to the continuum level. 
       The thin curve is the observed KPN spectrum, and the solid
       the CSII calculation
     with an assumed microturbulence $\xi_{\rm t} = 1$~km~s$^{-1}$.
     In this calculation no allowance for missing opacity has
     been made, and the continuum has been adopted as described.
     }
      \label{Ha4th.ps}
   \end{figure}

When spectra normalized to the continuum levels are compared,
we find an excellent agreement for H$_{\alpha}$ 
(Fig.~\ref{Ha4th.ps}). 
The results are the same both from the CSII and the SYNTHE
code, and are to be compared with BPO's Figure 8 (upper),
done for the solar flux.  We see good agreement in all cases.
The agreement of the CSII profiles with BPO profiles is expected, 
since the only basic difference 
is the use in BPO of the PK approximation while CSII
uses a full numerical convolution, a distinction we have found
thus far to be unimportant.  

\begin{figure}
\resizebox{7.0cm }{!}
{\includegraphics[angle=-90]{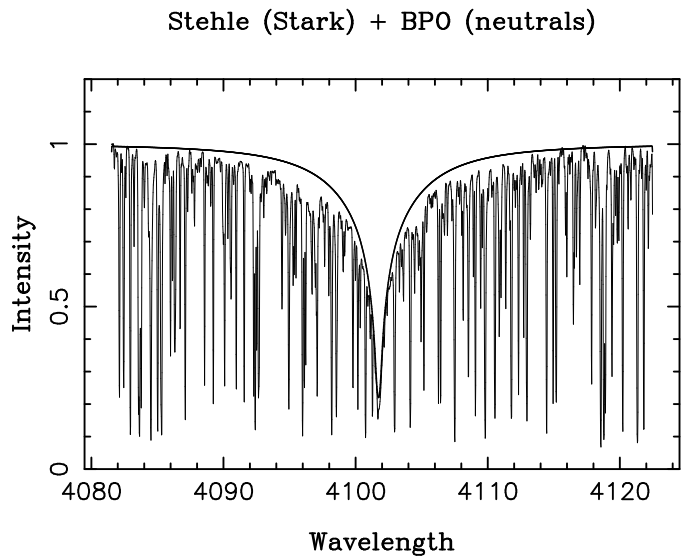}}
      \caption{KPN spectrum and CSII calculation for  H$\delta$.}  
      \label{Hd4th.ps}
   \end{figure}

As far as  the three higher, normalized Balmer lines are concerned, the
best fits to the wings are obtained when the ``observed'' continua
are adjusted downward from values obtained by
interpolation in Tab.~\ref{tab:cont}---the sense is that the continuum
there is too high.
For H$\gamma$ and H$\delta$, the downward adjustment is 2~\%.  The
observed continuum at H$\beta$ needed a downward adjustment of 3~\%;
problems with the continuum in this region were mentioned earlier
in this section.
Fig.~\ref{Hd4th.ps} shows the fit for H$\delta$.  The other two
Balmer line fits may be seen at the url:
http://www.astro.lsa.umich.edu/users/cowley/\-balmers.html/

In principle, the adjustment of the continuum
requires an iteration with a new continuous opacity
to the new continuum.  Fortunately, the
normalized Balmer profiles are not very sensitive to small adjustments
for the missing opacity.



\section{Inhomogeneities and the Plane Parallel Model}

For perhaps a century we have known that the spectrum of the 
solar photosphere varies from one point on the disk to another.  The first 
high-resolution spectra obtained from the McMath-Hulbert 
Observatory showed striking spatial variations that came to be 
known as ``wiggley lines.''  The solar line profiles vary 
markedly, both in time and space, and while we have understood the 
general the nature and cause of these variations for decades, 
recent numerical calculations by Nordlund, Stein, and their 
collaborators have provided a detailed description (cf. Nordlund
\& Stein 2001). 

In spite of its origin in a turbulent roil, the average line
spectrum of the sun is remarkably constant.  This is 
particularly surprising in the case of the Balmer lines, where
the large Boltzmann factor ($\theta\chi_{\rm lower} \approx 
10$) suggests huge local non-linear effects.  Naively, one 
would not expect them to average out, and the extent to which
they do average out remains to be fixed.  

In the 1950's, de Jager (1952) attempted to fix the 
temperature fluctuations in the solar atmosphere by making use 
of the putative nonlinearities of the Balmer lines.  His 
conclusions, of temperature differences of a thousand degrees
from hot to cool columns agrees remarkably with modern
numerical models.  Surely, he was guided by physical insight
into what the answer needed to be.  The Stark-broadening 
theory of that time was rudimentary, and the influence of 
collisions with neutral hydrogen were entirely neglected. 

We have found that reasonable matches to the four lower Balmer 
lines can be achieved using modern Stark profiles provided 
recent parameters for broadening by neutral hydrogen by BPO 
and the HM model are used.  In fact, the fits illustrated in
Figs.~\ref{Ha4th.ps} and ~\ref{Hd4th.ps}, were all based
on the empirical plane-parallel Holweger-M\"{u}ller model,
and include no attempts to improve the fits by plausible adjustments of 
the line-broadening parameters.  Other studies have explored 
the sensitivity of the Balmer lines to different theoretical
model atmospheres and to variations in the
convective mixing length to the pressure scale height ($l/H$).  

We remark here on the surprising {\it linearity} of the
Balmer profiles with the temperature of plane-parallel models.
This may be illustrated in several ways.  
In Fig.~\ref{wing4.ps} we can see that for $T_{\rm eff}$ about 
4000K to 6250K the wing strengths  
plot nearly linearly with temperature for the three higher
gravities. 

\begin{figure}
\resizebox{8.0cm}{!}
{\includegraphics[angle=-90]{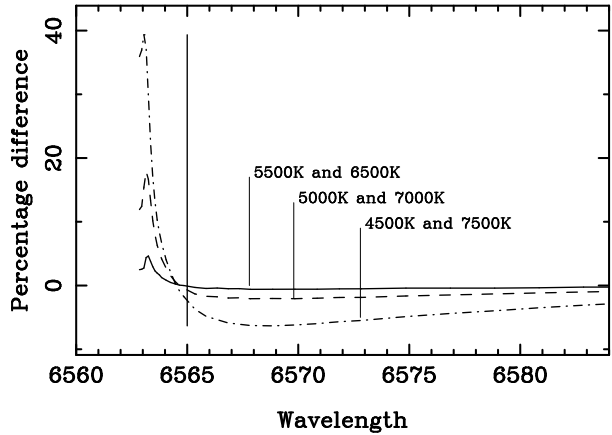}}
      \caption{Percentage differences in H$\alpha$ profiles
       for 6000K model and average profiles for three pairs of
       models as indicated (H$_{\alpha}$ profiles from Kurucz 1993a).}
      \label{wngdfs.ps}
   \end{figure}

This near linearity holds for most points on the line profiles,
apart from the most central portions.  
If one takes an equally
weighted average of H$\alpha$ fluxes for T = 5500K and 6500K,
the resulting mean differs imperceptibly from that for a
6000K model.  Means for a 5000K and 7000K model differ only
by 2\% from the 6000K model beyond 3A from the line center.
Even for the mean of a 4500K and an 7500K model is the
difference of the order of 5\% (see Fig.~\ref{wngdfs.ps}).

The same effect may be seen in the left panel of Fig. 3 
of Fuhrmann et al. (1993).   They show a series of 
Balmer profiles from H$\alpha$ through H$\delta$
for \logg=4, with effective temperatures running from 5000K
to 6700K, in steps of 100K.  It can be seen that the different
profiles are, for the most part, quite evenly spaced.   

The simple means of Fig.~\ref{wngdfs.ps} are certainly not 
equivalent to the detailed calculation performed, for example, by 
Asplund, Nordlund, \& Trampedach (1999), based on
the 3-dimensional numerical models of the solar convection
zone.  Nevertheless, they demonstrate that the non-linearities
that one might expect from the very large Boltzmann factors
of the $n = 2$ level are not realized in the {\it resultant}
Balmer profiles of cool stars.  This, in turn, supports endeavors
to use theoretical profiles from simplified stellar models to
help fix fundamental stellar parameters.   

\section{Conclusions}

We have explored recent techniques for computing Balmer line profiles
in the sun, and H$\alpha$ profiles in several models with
temperatures ranging from 4500K to 12000K.  We find that new Stark
profiles, rigorous convolution, and  improved  interpolation techniques
make almost no difference in the resulting calculated profiles,
compared with algorithms used in the Kurucz codes for several decades.

Good fits to normalized disk center solar profiles for the H$\alpha$ through
H$\delta$ are obtained from the Holweger-M\"{u}ller (HM) model.  The
H$\alpha$ profile can also be reasonably fitted in absolute intensity, but
the calculated continua for H$\beta$ through H$\delta$ are too high.
This may reasonably be attributed to missing UV opacity, perhaps
also to inadequacies of the HM model used here, as well as to
uncertainties in the absolute solar calibration.

In spite of severe temperature inhomogeneities in the solar atmosphere,
the plane-parallel model appears remarkably robust.

\section{Acknowledgement}
 
Numerous scientific colleagues have kindly consulted with us on
various parts of this project, and we will doubtless omit some
unintentionally.  For this we apologize.  Explicit thanks are due
to Drs. P. Barklem, N. Grevesse, K. Fuhrmann, R. L. Kurucz,
M. Lemke, H. Neckel, J. Sauval,
B. Smalley,  and C. Stehl\'{e}.

\appendix

\section{The Holweger-M\"uller (HM)  model as input for the Kurucz codes}
 
Table~A.1 lists the HM model
interpolated on 50 depths and converted to the 
RHOX scale of the Kurucz codes. The last four columns are the input model
for the Kurucz codes.  The units for the continuous
opacity $\kappa_{5000}$ are in cm$^{2}$~gr$^{-1}$. 

\begin{table*}  
\caption{The HM solar model}
\label{model} 
\begin{tabular}{rllllll} 
\hline
$\log\tau_{5000}$ &
\multicolumn{1}{c}{$\kappa_{5000}$} &
\multicolumn{1}{c}{$\rho$} &
\multicolumn{1}{c}{ RHOX} &
\multicolumn{1}{c}{ T} &
\multicolumn{1}{c}{P$_{gas}$}&
\multicolumn{1}{c}{N$_{e}$}\\
-6.54& 6.133~10$^{-4}$&4.956~10$^{-11}$& 4.7137~10$^{-4}$& 3900& 1.274~10$^{1}$&2.448~10$^{9}$ \\
-6.39& 6.606~10$^{-4}$&6.959~10$^{-11}$& 6.6340~10$^{-4}$& 3910& 1.791~10$^{1}$&3.394~10$^{9}$\\
-6.23& 7.200~10$^{-4}$&9.569~10$^{-11}$& 9.1603~10$^{-4}$& 3924& 2.472~10$^{1}$&4.605~10$^{9}$\\
-6.08& 7.927~10$^{-4}$&1.294~10$^{-10}$& 1.2441~10$^{-3}$& 3939& 3.357~10$^{1}$&6.146~10$^{9}$  \\
-5.93& 8.833~10$^{-4}$&1.723~10$^{-10}$& 1.6643~10$^{-3}$& 3960& 4.487~10$^{1}$&8.096~10$^{9}$ \\
-5.77& 9.899~10$^{-4}$&2.262~10$^{-10}$& 2.1990~10$^{-3}$& 3988& 5.943~10$^{1}$&1.055~10$^{10}$\\
-5.62& 1.119~10$^{-3}$&2.933~10$^{-10}$& 2.8759~10$^{-3}$& 4022& 7.762~10$^{1}$&1.366~10$^{10}$\\
-5.47& 1.278~10$^{-3}$&3.772~10$^{-10}$& 3.7234~10$^{-3}$& 4052& 1.007~10$^{2}$&1.747~10$^{10}$ \\
-5.31& 1.457~10$^{-3}$&4.085~10$^{-10}$& 4.7816~10$^{-3}$& 4084& 1.291~10$^{2}$&2.218~10$^{10}$\\
-5.16& 1.673~10$^{-3}$&6.072~10$^{-10}$& 6.0935~10$^{-3}$& 4120& 1.648~10$^{2}$&2.793~10$^{10}$\\
-5.01& 1.931~10$^{-3}$&7.617~10$^{-10}$& 7.7163~10$^{-3}$& 4159& 2.084~10$^{2}$&3.507~10$^{10}$\\
-4.85& 2.239~10$^{-3}$&9.523~10$^{-10}$& 9.7130~10$^{-3}$& 4188& 2.624~10$^{2}$&4.365~10$^{10}$\\
-4.70& 2.599~10$^{-3}$&1.183~10$^{-9}$&  1.2160~10$^{-2}$& 4220& 3.289~10$^{2}$&5.415~10$^{10}$ \\
-4.55& 3.022~10$^{-3}$&1.463~10$^{-9}$&  1.5156~10$^{-2}$& 4255& 4.102~10$^{2}$&6.684~10$^{10}$\\
-4.39& 3.528~10$^{-3}$&1.803~10$^{-9}$&  1.8818~10$^{-2}$& 4286& 5.093~10$^{2}$&8.221~10$^{10}$\\
-4.24& 4.118~10$^{-3}$&2.214~10$^{-9}$&  2.3278~10$^{-2}$& 4317& 6.295~10$^{2}$&1.006~10$^{11}$\\
-4.09& 4.814~10$^{-3}$&2.713~10$^{-9}$&  2.8713~10$^{-2}$& 4349& 7.762~10$^{2}$&1.229~10$^{11}$\\
-3.93& 5.630~10$^{-3}$&3.313~10$^{-9}$&  3.5330~10$^{-2}$& 4382& 9.572~10$^{2}$&1.497~10$^{11}$\\
-3.78& 6.572~10$^{-3}$&4.038~10$^{-9}$&  4.3379~10$^{-2}$& 4415& 1.175~10$^{3}$&1.824~10$^{11}$\\
-3.63& 7.703~10$^{-3}$&4.914~10$^{-9}$&  5.3175~10$^{-2}$& 4448& 1.439~10$^{3}$&2.212~10$^{11}$\\
-3.47& 9.024~10$^{-3}$&5.975~10$^{-9}$&  6.5074~10$^{-2}$& 4477& 1.762~10$^{3}$&2.679~10$^{11}$\\
-3.32& 1.059~10$^{-2}$&7.256~10$^{-9}$&  7.9518~10$^{-2}$& 4506& 2.153~10$^{3}$&3.237~10$^{11}$\\
-3.17& 1.241~10$^{-2}$&8.797~10$^{-9}$&  9.7037~10$^{-2}$& 4536& 2.630~10$^{3}$&3.911~10$^{11}$\\
-3.02& 1.453~10$^{-2}$&1.065~10$^{-8}$&  1.1833~10$^{-1}$& 4568& 3.206~10$^{3}$&4.723~10$^{11}$\\
-2.86& 1.706~10$^{-2}$&1.290~10$^{-8}$&  1.4414~10$^{-1}$& 4597& 3.899~10$^{3}$&5.682~10$^{11}$\\
-2.71& 2.000~10$^{-2}$&1.561~10$^{-8}$&  1.7550~10$^{-1}$& 4624& 4.753~10$^{3}$&6.822~10$^{11}$\\
-2.56& 2.347~10$^{-2}$&1.888~10$^{-8}$&  2.1346~10$^{-1}$& 4651& 5.781~10$^{3}$&8.192~10$^{11}$\\
-2.40& 2.749~10$^{-2}$&2.281~10$^{-8}$&  2.5958~10$^{-1}$& 4681& 7.031~10$^{3}$&9.854~10$^{11}$\\
-2.25& 3.221~10$^{-2}$&2.753~10$^{-8}$&  3.1560~10$^{-1}$& 4716& 8.551~10$^{3}$&1.184~10$^{12}$\\
-2.10& 3.776~10$^{-2}$&3.321~10$^{-8}$&  3.8363~10$^{-1}$& 4754& 1.040~10$^{4}$&1.429~10$^{12}$\\
-1.94& 4.418~10$^{-2}$&3.998~10$^{-8}$&  4.6626~10$^{-1}$& 4799& 1.262~10$^{4}$&1.729~10$^{12}$\\
-1.79& 5.171~10$^{-2}$&4.814~10$^{-8}$&  5.6680~10$^{-1}$& 4846& 1.535~10$^{4}$&2.092~10$^{12}$\\
-1.64& 6.053~10$^{-2}$&5.782~10$^{-8}$&  6.8889~10$^{-1}$& 4903& 1.866~10$^{4}$&2.544~10$^{12}$\\
-1.48& 7.103~10$^{-2}$&6.942~10$^{-8}$&  8.3714~10$^{-1}$& 4964& 2.270~10$^{4}$&3.105~10$^{12}$\\
-1.33& 8.318~10$^{-2}$&8.311~10$^{-8}$&  1.0172~10$^{0}$&  5040& 2.754~10$^{4}$&3.824~10$^{12}$\\
-1.18& 9.787~10$^{-2}$&9.934~10$^{-8}$&  1.2355~10$^{0}$&  5122& 3.350~10$^{4}$&4.726~10$^{12}$\\
-1.02& 1.157~10$^{-1}$&1.184~10$^{-7}$&  1.4988~10$^{0}$&  5217& 4.064~10$^{4}$&5.909~10$^{12}$\\
-0.87& 1.374~10$^{-1}$&1.410~10$^{-7}$&  1.8150~10$^{0}$&  5308& 4.920~10$^{4}$&7.396~10$^{12}$\\
-0.72& 1.651~10$^{-1}$&1.669~10$^{-7}$&  2.1921~10$^{0}$&  5416& 5.957~10$^{4}$&9.425~10$^{12}$\\
-0.56& 2.054~10$^{-1}$&1.950~10$^{-7}$&  2.6321~10$^{0}$&  5567& 7.145~10$^{4}$&1.263~10$^{13}$\\
-0.41& 2.756~10$^{-1}$&2.225~10$^{-7}$&  3.1174~10$^{0}$&  5781& 8.472~10$^{4}$&1.875~10$^{13}$\\
-0.26& 4.009~10$^{-1}$&2.470~10$^{-7}$&  3.6118~10$^{0}$&  6032& 9.817~10$^{4}$&3.037~10$^{13}$\\
-0.10& 6.179~10$^{-1}$&2.667~10$^{-7}$&  4.0810~10$^{0}$&  6315& 1.109~10$^{5}$&5.255~10$^{13}$\\
 0.05& 9.745~10$^{-1}$&2.812~10$^{-7}$&  4.5088~10$^{0}$&  6617& 1.227~10$^{5}$&9.274~10$^{13}$\\
 0.20& 1.471~10$^{0}$&2.933~10$^{-7}$&   4.9028~10$^{0}$&  6902& 1.334~10$^{5}$&1.545~10$^{14}$ \\
 0.35& 2.392~10$^{0}$&2.988~10$^{-7}$&   5.2624~10$^{0}$&  7266& 1.432~10$^{5}$&2.810~10$^{14}$\\
 0.51& 3.978~10$^{0}$&2.988~10$^{-7}$&   5.5724~10$^{0}$&  7679& 1.517~10$^{5}$&5.172~10$^{14}$\\
 0.66& 6.175~10$^{0}$&2.979~10$^{-7}$&   5.8448~10$^{0}$&  8059& 1.592~10$^{5}$&8.643~10$^{14}$\\
 0.81& 8.426~10$^{0}$&3.007~10$^{-7}$&   6.1102~10$^{0}$&  8335& 1.663~10$^{5}$&1.225~10$^{15}$\\
 0.97& 1.018~10$^{1}$&3.084~10$^{-7}$&   6.4090~10$^{0}$&  8500& 1.746~10$^{5}$&1.512~10$^{15}$\\
\hline

\end{tabular}
\end{table*}



\begin{thebibliography}{}

\bibitem[1965]{ag65} Ali,A.W., \& Griem,H.R.,1965, Phys.Rev., 140, 1044
\bibitem[1966]{ag66} Ali,A.W., \& Griem,H.R.,1965, Phys.Rev., 144, 366
\bibitem[1995]{am95} Anstee,S.D., \& O'Mara,B.J., 1995, MNRAS 276, 859
\bibitem[1999]{asp99} Asplund, M., Nordlund, \AA, \& Trampedach, R.
1999, in {\it Theory and tests of convection in stellar structure}, 
A. Gimenez, E. F. Guinan, and B. Montesinos (eds.), ASP Conference
Series, 173, 221
\bibitem[2000]{pbo00} Barklem,P.S., Piskunov,N., \& O'Mara,B.J., 2000, A\&A, 363, 1091 (BPO)
\bibitem[1995]{bur95} Burlov-Vasiljev,K.A., Gurtovenko,E.A., \& Matvejev,YU.B., 1995,
Solar Phys., 157, 51
\bibitem[1997]{gck97} Castelli,F., Gratton,R.G., \& Kurucz,R.L., 1997, A\&A, 
318, 841 (CGK)
\bibitem[1959]{dejag59} de Jager,C., 1959, in Handbuch der Physik, LII, p. 80, (Berlin:
   Springer). See Table 4, p. 105
\bibitem[1993]{fag93} Fuhrmann,K., Axer,M., \& Gehren,T., 1993, A\&A, 271, 451
\bibitem[1994]{fag94} Fuhrmann K., Axer,M., \& Gehren,T., 1994, A\&A, 285, 585
\bibitem[1999]{gks99} Gardiner,R.B., Kupka,F., \& Smalley,B., 1999, A\&A, 347, 876
\bibitem[1998]{gs98} Grevesse,N., \& Sauval,A.J., 1998, Space Sci. Rev., 85, 161
\bibitem[1974]{hm74} Holweger, H., \& M\"{u}ller, E. A., 1974, Sol. Phys., 39, 19.
\bibitem[1993]{k93a} Kurucz, R.L. 1993a, {\it ATLAS9 Stellar Atmosphere Programs and
2 km/sec grid}, CD-ROM No. 13 (Smithsonian Ap. Obs.)
\bibitem[1993]{k93b} Kurucz, R.L. 1993b, {\it SYNTHE Spectrum Synthesis
Programs and Line Data}, CD-ROM No. 18 (Smithsonian Ap. Obs.)
\bibitem[1993]{k93c} Kurucz, R.L. 1993c, {\it Opacities for Stellar
Atmospheres}, CD-ROM No. 2 (Smithsonian Ap. Obs.)
\bibitem[1969]{lr69} Lortet, M.C., \& Roueff, E., 1969, A\&A, 3, 462
\bibitem[1999]{nek99} Neckel, H., 1999, Solar Phys., 184, 421 (KPN)
\bibitem[1984]{nl84}  Neckel, H., \& Labs, D., 1984, Solar Phys., 90, 205
\bibitem[1994]{nl94}  Neckel, H., \& Labs, D., 1994, Solar Phys., 153, 91
\bibitem[2001]{nor01} Nordlund,\AA, \&  Stein,R.F., 2001, ApJ, 546, 576
\bibitem[1993]{pet93} Peterson,D. 1993, see documentation in 
{\it SYNTHE Spectrum 
   Synthesis Programs and Line Data}, Kurucz CD-Rom 18, 
   Smithsonian Ap. Obs.  See also programs and documentation on 
   the Kurucz web site: http://kurucz.harvard.edu
\bibitem[2001]{pet01} Peterson, R.C., Dorman, B., \& Rood, R.T., 2001,
ApJ, 559, 372
\bibitem[1992]{sea90} Seaton,M.J., Zeippen,C.J., Tully,J.A., Pradhan,A.K., 
Mendoza,C., Hibbert,A., \& Berrington,K.A., 1992, Rev. Mexicana
Astron. Astrofis., 23, 19 (see also
http://cdsweb.u-strasbg.fr/topbase.html)
\bibitem[1996]{sneden96} Sneden,C., McWilliam,A., Preston,G.W., Cowan,J.J., 
   Burris,D.B., \& Armosky,B.J., 1996, ApJ, 467, 819
\bibitem[1999]{sh99} Stehl\'{e},C., \& Hutcheon,R., 1999, A\&AS, 140, 93      
\bibitem[1955]{uns55} Uns\"{o}ld,A., 1955, {\it Physik der
Sternatmosph\"{a}ren,} 2nd ed. (Berlin: Springer)
\bibitem[1996]{vant96} van't Veer-Menneret,C., \& M\'{e}gessier,C., 1996, A\&A, 309, 879
\bibitem[1973]{vcs73} Vidal,C. R., Cooper,J., \& Smith,E.W., 1973, ApJS, 25, 37 (VCS)
\end{thebibliography}
\end{document}